\documentclass[structabstract]{aa}

\usepackage{graphicx}
\usepackage{txfonts}
\usepackage{natbib}
\usepackage{array}
\bibpunct{(}{)}{;}{a}{,}{,}

\begin{document}
\bibliographystyle{aa}

\title{Exploring the Nuclear Environment of the NLS1 Galaxy \\
 Arakelian 564 with XMM-Newton RGS}
\titlerunning{XMM-Newton RGS analysis of Arakelian 564}
\author{R.A.N. Smith\thanks{rans@mssl.ucl.ac.uk} \and M.J. Page \and G. Branduardi-Raymont}
\authorrunning{R.Smith, M.Page and G.Branduardi-Raymont}
\date{\today}
\institute{Mullard Space Science Laboratory, University College London, Holmbury St. Mary, Dorking, Surrey RH5 6NT, England}

\abstract
{}
{We present an accurate characterisation of the high-resolution X-ray spectrum of the Narrow Line Seyfert 1 galaxy Arakelian 564 and put it in to context with other objects of its type by making a detailed comparison of their spectra.}
{The data are taken from 5 observations with the \textit{XMM-Newton} Reflection Grating Spectrometer and fitted with various spectral models.}
{The best fit to the data identifies five significant emission lines at 18.9, 22.1, 24.7, 29.0 and 33.5\AA~due O\,VIII Ly$\alpha$, O\,VII(f), N\,VII Ly$\alpha$, N\,VI(i) and C\,VI Ly$\alpha$ respectively. These have an RMS velocity of $\sim1100$\,km\,s$^{-1}$ and a flow velocity of $\sim-600$\,km\,s$^{-1}$, except for the O\,VII(f) emission line, which has a flow velocity consistent with zero. Two separate emitting regions are identified. Three separate phases of photoionized, X-ray absorbing gas are included in the fit with ionization parameters log\,$\xi=-0.86$, 0.87, 2.56 and column densities $N_H=0.89, 2.41, 6.03\times10^{20}$\,cm$^{-2}$ respectively. All three phases show this to be an unusually low velocity outflow ($-10\pm100$\,km\,s$^{-1}$) for a narrow line Seyfert 1. We present the hypothesis that the BLR is the source of the NLR and warm absorber, and examine optical and UV images from the \textit{XMM-Newton} Optical Monitor to relate our findings to the characteristics of the host galaxy.}
{}


\maketitle

\section{Introduction}
\label{564intro}

Narrow Line Seyfert 1 galaxies (NLS1s) were initially classified by \cite{osterbrock85} and are defined as having H$\beta$ FWHM\,$\leq$\,2000\,km\,s$^{-1}$. Their X-ray spectra often display more rapid variability (e.g.~\nocite{boller96}Boller et al. 1996;~\nocite{laor94}Laor et al. 1994) and are steeper in the soft X-ray band than those of `normal' broad line Seyfert 1s (e.g.~\nocite{leighly99}Leighly 1999). From the \textit{ROSAT} All Sky Survey it was found that approximately half of the sources in  soft X-ray selected samples of AGN are NLS1s (\nocite{grupe96}Grupe 1996;~\nocite{hasinger97}Hasinger 1997). They are believed to be `high-state' active galaxies, with low black hole masses for their luminosity, and so with high accretion rates relative to Eddington (\nocite{pounds95}Pounds et al. 1995;~\nocite{boroson02}Boroson 2002).
\newline

Observations indicate that the majority of Seyfert 1s display evidence for `warm absorbers' (\nocite{blustin05}Blustin et al. 2005), i.e. clouds of photoionized X-ray absorbing gas in our line of sight (e.g.~\nocite{komossa00}Komossa 2000), yet still very little is known of the absorbers origin, location or structure. The warm absorber is often characterised by its ionization parameter, $\xi$. This is a measure of the ionization state of the material and is defined as

\begin{equation}
\xi=\frac{L}{nr^2}
\label{564_xi_eq}
\end{equation}
where $L$ is the source luminosity (in erg\,s$^{-1}$), $n$ is the gas number density (in cm$^{-3} $) and $r$ is the distance of the absorber from the central engine (in cm) (\nocite{tarter69}Tarter et al. 1969). 
\newline

Arakelian 564 is the brightest known NLS1 in the 2--10\,keV range (L$_{2-10\,keV}=2.4\times10^{43}\,$erg\,s$^{-1}$, \nocite{turner01} Turner et al. 2001). This, coupled with its close proximity, z=0.02468 (\nocite{huchra99}Huchra et al. 1999), and relatively low Galactic column density, N$_H=5.34\times10^{20}$cm$^{-2}$ (\nocite{kalberla05}Kalberla et al. 2005), make it a very interesting target to investigate. This object has been studied across all wavebands (e.g.~\nocite{shemmer01}Shemmer et al. 2001;~\nocite{romano04}Romano et al. 2004). Here we have merged data from 5 observations with the \textit{XMM-Newton} Reflection Grating Spectrometer (RGS), allowing us to explore, in unprecedented detail, the high-resolution X-ray spectrum of this object.
\newline

\citet{matsumoto04} report on a 50\,ks \textit{Chandra} HETGS observation. They fit the hard X-ray spectrum with a power law of photon index of $2.56\pm0.06$ and add a blackbody, of temperature $0.124\pm0.003$\,keV, to fit the soft excess. They detect an edge-like feature at 0.712\,keV, and two phases of absorption by photoionized gas (phase 1: log\,$\xi\sim1$, N$_{H}\sim10^{21}$\,cm$^{-2}$; phase 2: log\,$\xi\sim2$, N$_{H}\sim10^{21}$\,cm$^{-2}$).~\citet{vignali04} have analysed two \textit{XMM-Newton} observations from June 2000 and June 2001. They fit the spectrum with a steep power law ($\Gamma=2.50-2.55$) plus a soft blackbody component (kT$\sim$140\,--\,150\,eV). They also identify an edge-like feature in the EPIC data at $\sim0.73$\,keV, which they interpret as an O\,VII K absorption edge, and find evidence for a broad iron emission line, at $\sim1$\,keV, in one observation. ~\citet{crenshaw02} have analysed the UV spectrum using data from the STIS onboard the \textit{Hubble Space Telescope}. They identify UV absorption lines due to Ly$\alpha$, N\,V, C\,IV, Si\,IV and Si\,III centred at a radial velocity of $-190$\,km\,s$^{-1}$ relative to the systemic velocity. They identify this with a dusty lukewarm absorber with column density N$_{H}=1.62\times10^{21}$\,cm$^{-2}$ and ionization U=0.033 (this ionization is equivalent to log\,$\xi\simeq1.2$, converted using the method of \nocite{george98}George et al. 1998). 
\newline

\citet{papadakis06} analysed the most recent (2005), and longest ($\sim100$\,ks), observation with \textit{XMM-Newton}, focusing on the EPIC data. They find that the soft excess cannot be parametrized either by a multiple power law or by a power law plus a single blackbody, in contrast to previous findings. They fit the MOS and PN spectra with either a power law of slope $2.43\pm0.03$ and two blackbodies, with kT$\sim$0.15 and $\sim$0.07\,keV, or with a relativistically blurred photoionized disk reflection model. They also find evidence for two phases of photoionized X-ray absorbing gas with ionization parameters log\,$\xi\sim1$ and $\sim2$ and column densities N$_{H}\sim2$ and $5\times 10^{20}$\,cm$^{-2}$.
\newline

The same 2005 data were also analysed by~\citet{dewangan07}. They concentrate their analysis on the EPIC data, but also study the RGS spectrum and find two warm absorber phases with ionization parameters log\,$\xi\sim2$ and $<0.3$, column densities $N_H\sim4\times10^{20}$ and $2\times10^{20}$cm$^{-2}$ and flow velocities $v\sim-300$ and $-1000$\,km\,s$^{-1}$ respectively. \newline

Here we investigate the combined RGS spectrum of Arakelian 564. Section~\ref{obs_data_red} gives information on the individual observations and describes our data reduction process. The spectral fitting and best fit results are presented in Section~\ref{analysis}. A full discussion of these results and comparison with other observations is given in Section~\ref{discussion}.

\section{Observations and Data Reduction}
\label{obs_data_red}

\begin{figure}
\includegraphics[angle=270, scale=0.37]{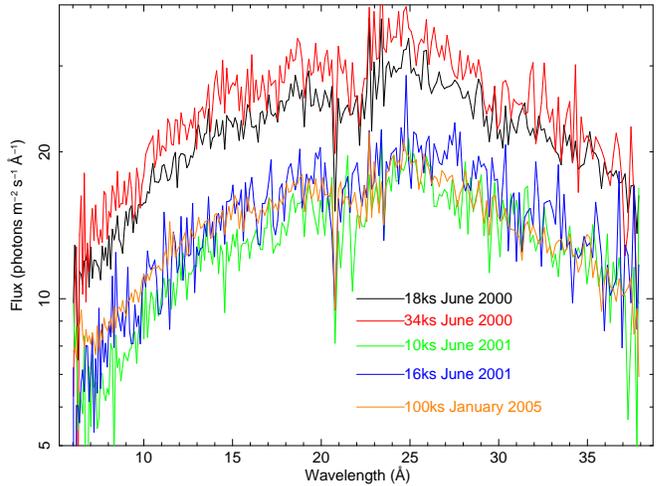}
\caption{All five \textit{XMM-Newton} RGS observations of Arakelian 564 plotted separately. A similar spectral shape is seen for all observations. The two high luminosity spectra are from the 2000 observations, the three lower are the two 2001 and one 2005 observations.}
\label{564_separate}
\end{figure}

The majority of our data were obtained with the Reflection Grating Spectrometer (\nocite{denherder01}den Herder et al. 2001) onboard the \textit{XMM-Newton} observatory. This spectrometer provides unparalleled sensitivity in the 6\,--\,38\AA~(0.3\,--\,2.5\,ke\,V) range with an approximately constant wavelength resolution of 70\,m\AA~FWHM. The data presented comprise four short exposure (18, 34, 10, 16\,ks) observations, two in June 2000 and two in June 2001 respectively, and one long exposure ($\sim$100\,ks) observation in January 2005, the EPIC data of which have been presented by \citet{papadakis06}. The overall appearance of the continuum shape is not significantly different between observations; as can be seen in Fig.~\ref{564_separate} there is a flux variation of around 50\% between observations. The X-ray flux is known to vary by $>$50\% within individual observations (e.g.~\nocite{papadakis06}Papadakis et al. 2007;~\nocite{vaughan99}Vaughan et al. 1999), hence the variations between our observations are relatively unimportant. There are also no significant changes in the warm absorber features between observations as we shall demonstrate in Section~\ref{564_indiv}. Therefore, data from all five \textit{XMM-Newton} observations can be combined to give the best possible signal-to-noise spectrum.
\newline

\begin{figure*}
\begin{center}
\includegraphics[angle=270, scale=0.6]{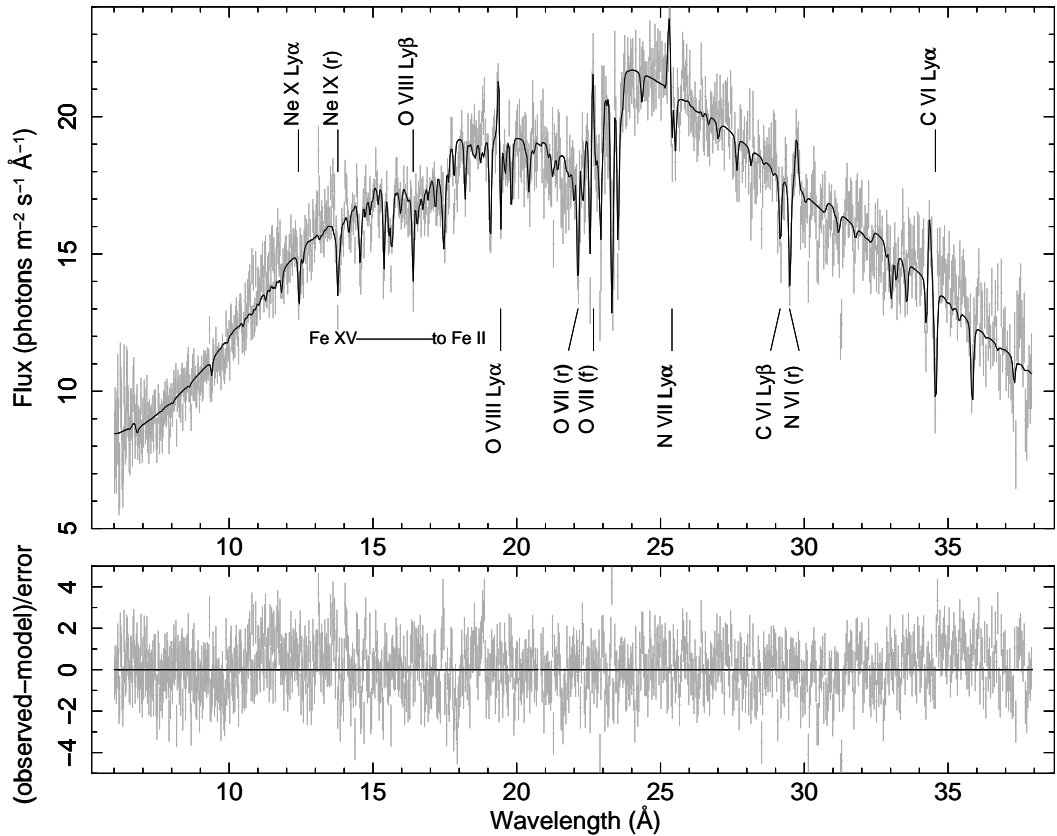}
\caption{The combined Arakelian 564 data from \textit{XMM-Newton} RGS fitted with a power law, blackbody, Galactic absorption, 3 phases of photoionized absorbing gas and five narrow emission lines due to O\,VIII Ly$\alpha$, O\,VII(f), N\,VII Ly$\alpha$, N\,VI(i) and C\,VI Ly$\alpha$. The bottom panel shows the residuals for this fit.}
\label{564_res}
\end{center}
\end{figure*}

The data were processed using the \textit{XMM-Newton} Science Analysis System (SAS) version 7.0. First and second order spectra from both RGS, and all observations, were extracted. The spectra and response matrices were resampled and coadded to produce a single spectrum (see Fig.~\ref{564_res}) and a single response matrix as described in~\cite{page03a}. To improve signal to noise the data were grouped by a factor of 3, resulting in a spectrum with $1000$ channels $\sim$\,30\,m\AA\ wide, well sampled with respect to the RGS resolution of $\sim$\,70\,m\AA\ FWHM. The main software used to model the data was \sc spex \rm 2.00, a spectral fitting package created by J. S. Kaastra, which is optimised for use with high resolution X-ray spectra. The data presented in the figures have not been corrected for redshift or Galactic absorption unless specified otherwise.
\newline

In addition to the RGS spectra we also investigate the nature of the host galaxy using archive B-band and I-band images, from 1.2\,ks exposures in November 1998 with the 1m Nickel Telescope at the Lick Observatory (\nocite{schmitt00}Schmitt \& Kinney 2000), and combined UV images, from the \textit{XMM-Newton} Optical Monitor (XMM-OM,~\nocite{mason01}Mason et al. 2001) taken simultaneously with the RGS observations mentioned above. The pipeline processed XMM-OM UV sky-images in each filter were aligned and then coadded to produce the image of Arakelian 564 for each of the UVW1 (220-400\,nm), UWM2 (200-280\,nm) and UVW2 (180-260\,nm) filters.

\begin{figure}
\includegraphics[angle=270, scale=0.37]{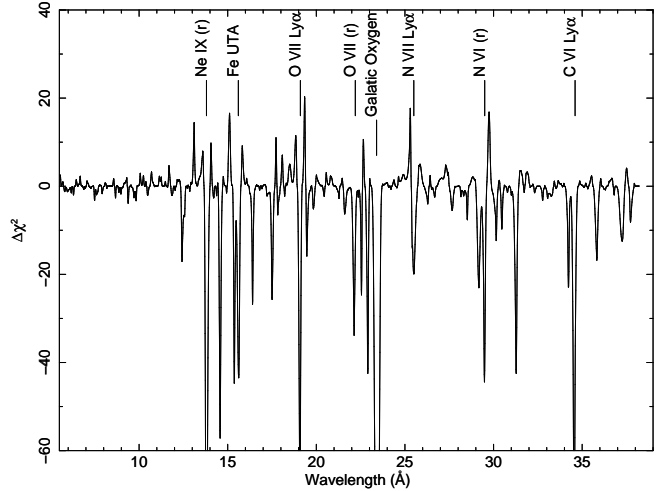}
\caption{A graph of $\Delta\chi^2$ against wavelength to show the significant spectral lines. A $\Delta \chi^2$ of 9 implies a 3$\sigma$ significance. The positive values correspond to emission features and the negative to absorption features. The strongest absorption lines are identified.}
\label{564_lines}
\end{figure}

\begin{figure*}
\begin{center}
\includegraphics[angle=270, scale=0.7]{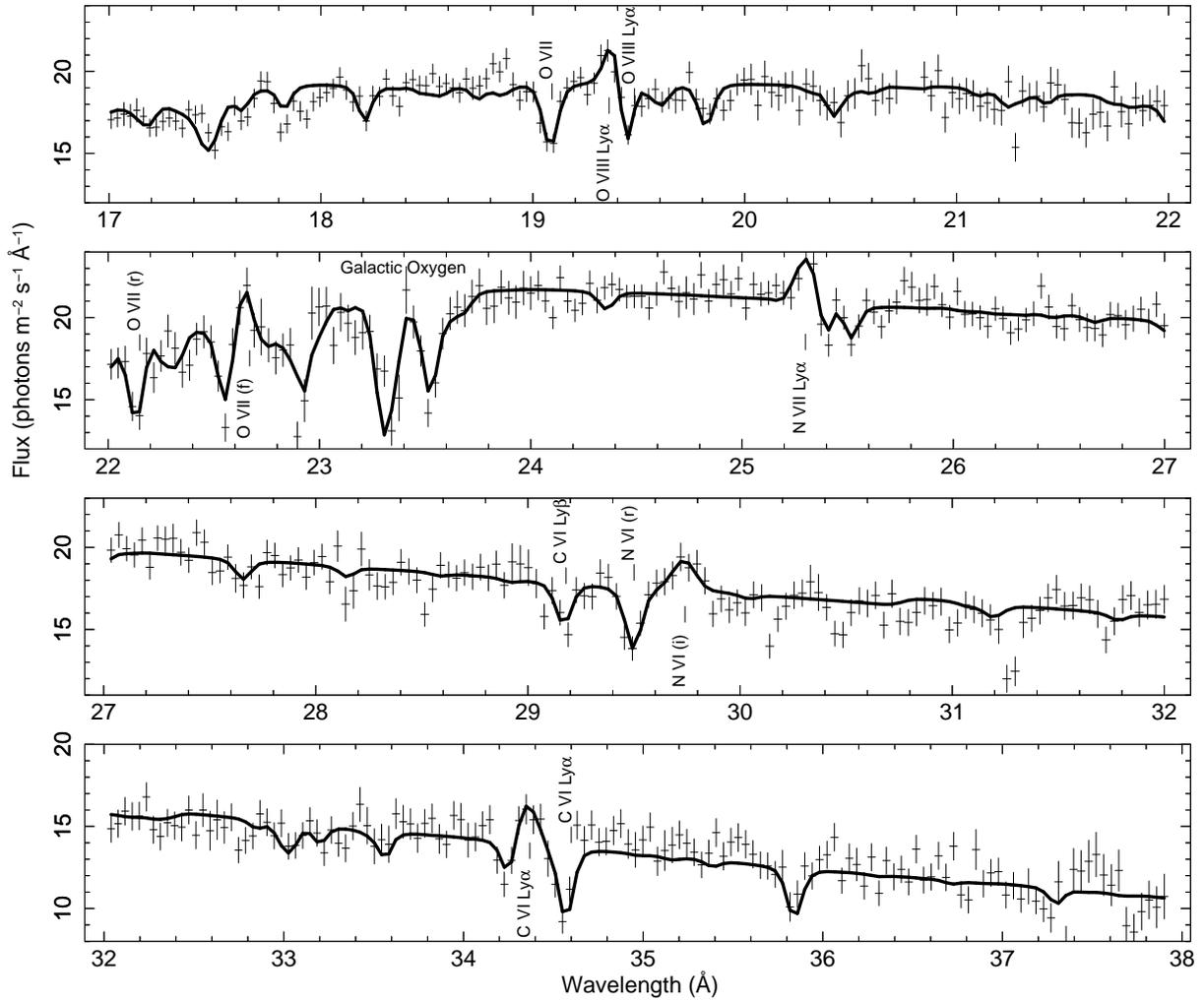}
\caption{A zoomed in section of the data to show the most prominent absorption and emission lines.}
\label{564_parts}
\end{center}
\end{figure*}

\section{Data Analysis}
\label{analysis}
\subsection{Continuum and Emission Features}

In order to correctly identify absorption and emission features a line-finding program (\nocite{page03a}Page et al. 2003) is used to detect significant absorption and emission lines in the spectrum. The program uses a sliding cell ($\sim1$\AA~wide) detection technique, rejecting all points more than $2\sigma$ deviant to create a smoothed continuum. A grid search is then performed, adding a Gaussian to the continuum model at each point and calculating the goodness of fit by minimising $\chi^2$ (\nocite{lampton76}Lampton et al. 1976). This produces a graph of $\Delta\chi^2$ against wavelength as shown in Fig.~\ref{564_lines}. A $\Delta\chi^2$ value of 9 corresponds to a $3\sigma$ significance and so, with the Arakelian 564 data, only one spurious statistical fluctuation would be expected at this significance level.
\newline

To begin the fitting process our continuum parameters are fixed at those required by the \citet{papadakis06} power law ($\Gamma=2.43$) and two blackbody (kT=150, 70\,eV) fit. We also include Galactic absorption ($N_H=5.34\times 10^{20}\,$cm$^{-2}$,~\nocite{kalberla05}Kalberla et al. 2005). 
\newline

Using Fig.~\ref{564_lines} as an initial guide, we can fit a number of emission lines to the data. The strongest three emission lines at $\sim19.40$, 25.35~and 29.75\AA~are well modelled with Gaussian profiles, and match the wavelengths of O\,VIII Ly$\alpha$, N\,VII Ly$\alpha$ and N\,VI(i) respectively. Using the Gaussian component in \sc spex \rm these emission lines are added to the above continuum and the profiles fitted to the data. The fit can be seen in Fig.~\ref{564_parts} on an expanded scale. Also included in the fit are emission lines from O\,VII(f) and C\,VI Ly$\alpha$ at $\sim22.65$ and 34.35\AA~respectively, which are required by the data (improvement in $\chi^2$ of $\sim50$ for 6 degrees of freedom: normalisation, wavelength and FWHM for each line) but are not as significant in Fig.~\ref{564_lines} due to the close proximity of absorption features. The parameters for all emission lines can be seen in Table~\ref{564_emis}, which lists the identification of each emission line, FWHM, flux and velocity characteristics.

\subsection{Absorption Features}

\setlength{\extrarowheight}{0.1cm}
\begin{table*}
\begin{center}
\caption{Table of parameters for the five narrow emission lines included in the fit (errors are 90\% confidence).}
\begin{tabular}{lcccccc}
\hline
\hline
Line & Observed $\lambda$ & Rest frame $\lambda$ & FWHM & FWHM & Flow Velocity & Line Luminosity \\
 & \AA & \AA & \AA & km\,s$^{-1}$ & km\,s$^{-1}$ & $10^{50}$\,ph\,s$^{-1}$\\
\hline
O\,VIII Ly$\alpha$ & $19.41^{+0.01}_{-0.01}$ & $18.94^{+0.01}_{-0.01}$ & $0.09^{+0.03}_{-0.02}$ & $1400^{+400}_{-350}$ & $-440^{+200}_{-200}$ & $2.11^{+0.98}_{-0.78}$ \\[3pt]
O\,VII (f) & $22.65^{+0.08}_{-0.05}$ & $22.10^{+0.08}_{-0.05}$ & $0.22^{+0.10}_{-0.22}$ & $3000^{+1500}_{-3000}$ & $90^{+1000}_{-700}$ & $3.10^{+1.82}_{-1.84}$ \\[3pt]
N\,VII Ly$\alpha$ & $25.34^{+0.02}_{-0.02}$ & $24.73^{+0.02}_{-0.02}$ & $0.12^{+0.07}_{-0.11}$ & $1500^{+800}_{-1400}$ & $-650^{+200}_{-250}$ & $1.75^{+0.47}_{-0.70}$ \\[3pt]
N\,VI (i) & $29.74^{+0.04}_{-0.05}$ & $29.02^{+0.04}_{-0.05}$ & $0.08^{+0.12}_{-0.08}$ & $850^{+1200}_{-850}$ & $-630^{+450}_{-550}$ & $0.88^{+0.51}_{-0.47}$ \\[3pt]
C\,VI Ly$\alpha$ & $34.36^{+0.02}_{-0.02}$ & $33.53^{+0.02}_{-0.02}$ & $0.10^{+0.08}_{-0.07}$ & $850^{+700}_{-600}$ & $-1800^{+200}_{-200}$ & $2.26^{+0.61}_{-0.55}$ \\[3pt]
\hline
\end{tabular}
\label{564_emis}
\end{center}
\end{table*}

\setlength{\extrarowheight}{0.1cm}
\begin{table*}
\begin{center}
\caption{Table of parameters for the three phases of photoionized, X-ray absorbing gas (errors are 90\% confidence).}
\begin{tabular}{lcccc}
\hline
\hline
Phase & Flow Velocity & RMS Velocity & Log\,$\xi$ & Column Density \\
 & (km\,s$^{-1}$) & (km\,s$^{-1}$) & (erg\,cm\,s$^{-1}$) & ($10^{20}$cm$^{-2}$) \\
\hline
Phase 1 & $20^{+200}_{-100}$ & $100^{+350}_{-40}$ & $-0.86^{+0.16}_{-0.17}$ & $0.89^{+0.34}_{-0.27}$ \\[3pt]
Phase 2 & $-40^{+50}_{-90}$ & $60^{+20}_{-10}$ & $0.87^{+0.06}_{-0.14}$ & $2.41^{+0.34}_{-0.34}$ \\[3pt]
Phase 3 & $-10^{+40}_{-100}$ & $100^{+30}_{-20}$ & $2.56^{+0.04}_{-0.04}$ & $6.03^{+0.97}_{-0.86}$ \\[3pt]
\hline
\end{tabular}
\label{564_xabs}
\end{center}
\end{table*}

The most significant absorption lines in Fig.~\ref{564_lines} match the wavelengths of Ne\,IX, O\,VIII, Galactic O\,I and O\,II, N\,VII and C\,VI. There are a number of spectral components in \sc spex \rm that can now be included in the fit to characterise these absorption lines. Here we use the \sc xabs \rm component. This is a self-consistent model, which contains all of the ions present in a photoionized gas at a particular ionization level. Free parameters in the fit are the ionization parameter, column density, RMS velocity and flow velocity; the abundances have been set at the Solar values (\nocite{verner96}Verner et al. 1996). Multiple `phases' of gas can be added to the model, using multiple \sc xabs \rm components, to reproduce the observed spectrum. We used the Arakelian 564 spectral energy distribution (SED) from~\citet{romano04} (this SED is consistent within errors with the OM and RGS average fluxes measured from our observations) to model the ionizing flux and compute the absorption profile in \sc spex\rm.
\newline

\begin{figure*}
\begin{center}
\includegraphics[angle=270, scale=0.6]{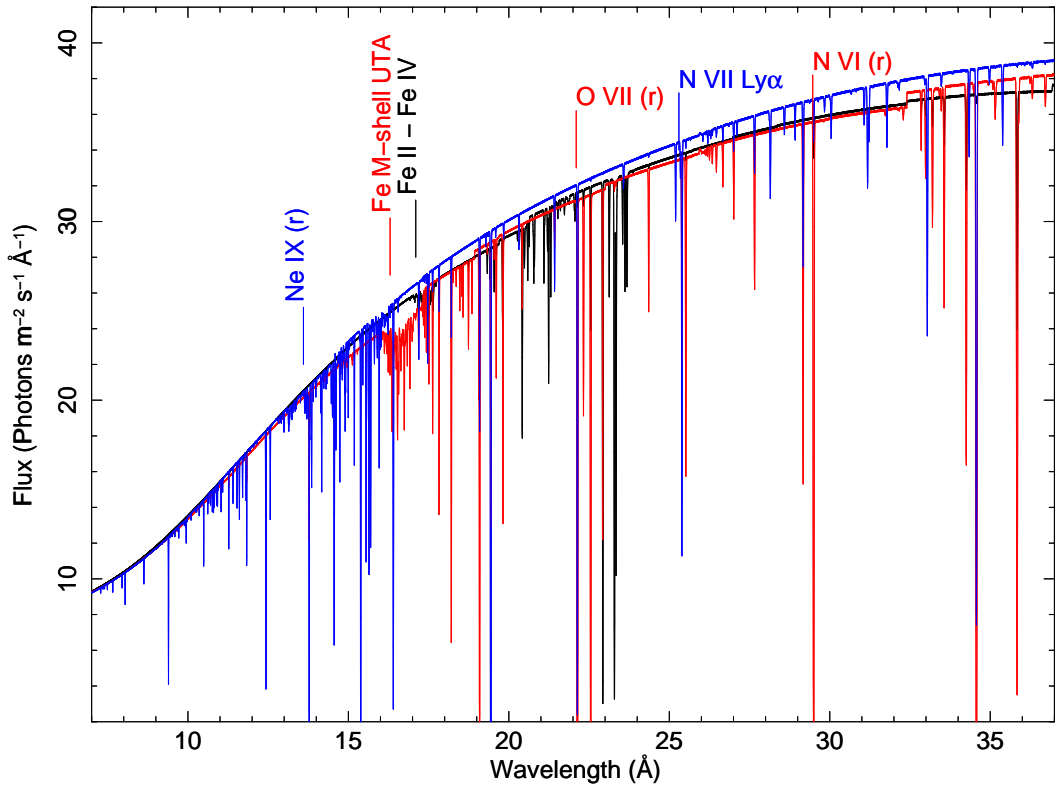}
\caption{The model fitted to the Arakelian 564 data, including three phases of photoionized, X-ray absorbing gas (Phase 1 in black, Phase 2 in red, Phase 3 in blue).}
\label{564_xmodel}
\end{center}
\end{figure*}

The best fit to the data includes three phases of photoionized gas, the model profile of which can be seen in Fig.~\ref{564_xmodel}; the fitted spectrum is in Fig.~\ref{564_res}. The parameters of the model are listed in Table~\ref{564_xabs}: each phase has a different ionization parameter and column density. All three phases have compatible velocity profiles; however, the flow velocities are not compatible with those of the emission lines (see Table~\ref{564_emis}), which are faster outflowing, with the exception of the O\,VII(f) line. The uncertainty in the wavelength scale of RGS data is $\sim8\,$m\AA, giving rise to a systematic uncertainty on the outflow velocities of at most 100\,km\,s$^{-1}$. Therefore, it is unlikely that the absorber and the majority of the emission lines are connected. However, Phases 2 and 3 are compatible with the O\,VII(f) emission line in ionization and velocity, so it is possible that these are connected. 

\subsection{Ionization structure}
\label{sec_ion}

It is possible that a continuous distribution of gas with a range of column densities and ionizations, rather than three discrete phases, is what is observed in Arakelian 564. To investigate this possibility further the data were fitted with the same continuum and emission lines as above, but using the \sc warm \rm component in place of the three \sc xabs \rm components. \sc warm \rm uses the equivalent of 19 \sc xabs \rm phases by default, each having an incremental logarithmic increase in ionization ranging between two user set ionization parameters. The column density is a free parameter for each of the 19 phases, but all have the same velocity profile.
\newline

\begin{figure}
\includegraphics[angle=270, scale=0.37]{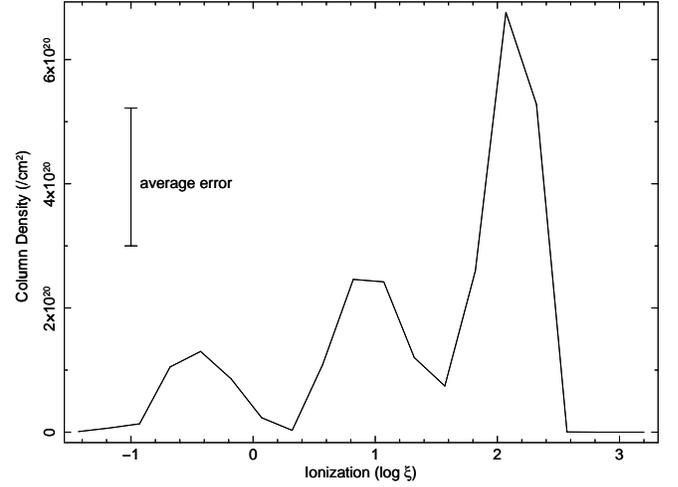}
\caption{The distribution of column density versus ionization parameter using the \sc warm \rm component in \sc spex \rm to fit the data with a continuous phase of gas.}
\label{564_warm_data}
\end{figure}

The result is similar to that using \sc xabs\rm. The best fit gives an RMS velocity value of $70^{+5}_{-5}$\,km\,s$^{-1}$, and flow velocity of $-40^{+35}_{-35}$\,km\,s$^{-1}$, which are compatible with the results from the \sc xabs \rm fit. Fig.~\ref{564_warm_data} shows the column density plotted against the ionization parameter for the \sc warm \rm component. Three peaks are clear in the data, at approximately the same ionization values and column densities as the three \sc xabs \rm phases fitted previously. This adds to the credibility of the initial fit and suggests that there are indeed three discrete phases of photoionized gas, rather than a continuous distribution of ionization parameter.
\newline

To investigate the possibility of all three phases co-existing in the same location we use the pressure form of the ionization parameter, $\Xi$. This is related to $\xi$ by

\begin{equation}
\Xi=\frac{L_{ion}}{4\pi cr^2 P}=\frac{0.961\times 10^4 \xi}{T}
\end{equation}
where $L_{ion}$ is the 1--1000\,Ryd ionizing luminosity, $r$ is the distance of the absorber from the ionizing source, $P$ is the pressure and $T$ the temperature (\nocite{krolik81}Krolik et al. 1981).
\newline

\begin{figure}
\includegraphics[angle=270, scale=0.37]{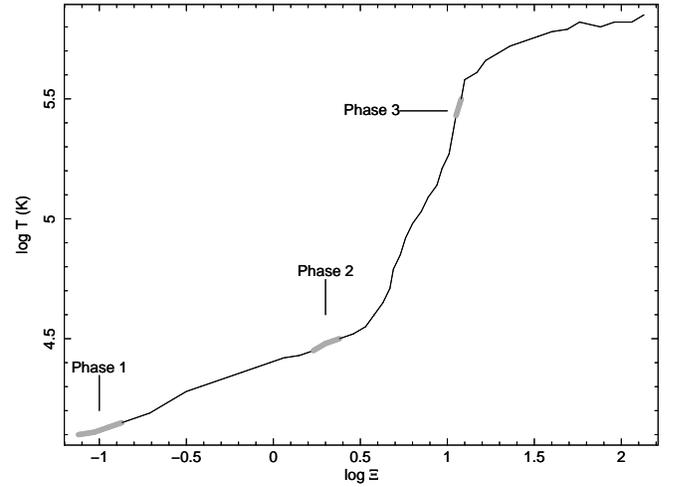}
\caption{Thermal stability graph showing the equilibrium temperature as a function of $\Xi$ for the SED described in~\citet{romano04}. The positions relating to the three warm absorber phases, along with their errors, are indicated.}
\label{564_preseq}
\end{figure}

The SED from~\citet{romano04} was used to create a graph of thermal stability of $\Xi$ against temperature, that is shown in Fig.~\ref{564_preseq}. The best fit ionization parameters and their errors, from Table~\ref{564_xabs}, are used to mark the regions corresponding to the three phases on the graph. In order for the phases to be in pressure equilibrium, and so co-exist, they must have overlapping values of $\Xi$. In Fig.~\ref{564_preseq} it can be seen that the three phases do not overlap in $\Xi$ and so cannot co-exist in pressure equilibrium. The majority of AGN display an unstable region where $d(lnT)/d(ln\Xi)<0$, as noted in~\citet{krolik81}. The lack of any unstable region is unusual, but the shape of the curve is very similar to that seen in the NLS1 IZw1 as presented by~\citet{costantini07}.
\newline

Constraints can be put on the location of the warm absorber phases. We can estimate the maximum distance that the gas can be from the continuum source, and still be ionized at the measured levels. We can assume that the thickness of any gas phase, $\Delta r$, has to be less than or equal to the distance from the source, $R$:

\begin{equation}
\frac{\Delta r}{R}\leq 1.
\end{equation} 
The column density, $N_H$, as a function of density, $n(R)$, is given by

\begin{equation}
N_H\sim n(R)C_v\Delta r
\end{equation}
where $C_v$ is the volume filling factor. This, together with Equation~\ref{564_xi_eq} gives the expression for the maximum distance

\begin{equation}
R\leq \frac{L_{ion}C_v(R)}{\xi N_H}
\label{564_Rleq}
\end{equation} 
(\nocite{blustin05}Blustin et. al 2005). We can estimate $L_{ion}$ by integrating over the SED from~\citet{romano04}: Log\,$L_{ion}=45.2$. We use a typical value of $C_v(R)$ of $0.03$ (from the average $C_v(R)$ of NLS1s calculated by~\nocite{blustin05}Blustin et al. 2005), since $C_v(R)$ cannot be calculated for Arakelian 564 because of its small outflow velocity. Values for the ionization parameter and column density are taken from Table~\ref{564_xabs}. This gives values of $R\leq$\,1000\,kpc, 10\,kpc and 100\,pc for Phases 1, 2 and 3 respectively. 
\newline

Additional information can be gained from looking at the UV spectra presented in~\citet{crenshaw02} and~\citet{romano02}. It is clear from the UV absorption and emission line profiles that the absorption eats away the emission lines and so the absorption occurs between us and the emission line region. Therefore, we can use the distance of the UV emission line region as a minimum distance to our warm absorber for Phase 1 and 2, since these are seen in the UV spectrum.~\citet{romano04} calculated the size of the $H\beta$ emitting line region, using the FWHM of UV and optical emission lines, as $R_{BLR}\simeq 0.008\pm0.002$\,pc.
\newline

\citet{crenshaw02} find that the UV absorber must lie outside the NLR and use the Ly$\alpha$ spatial profile to put a limit on the size of the NLR of $\leq95$\,pc. Since Phases 1 and 2 have been shown to contribute significantly to the UV spectrum they must also lie outside this region.
\newline

We can also estimate the minimum distance to the absorber by assuming that the outflow velocity has to be greater than or equal to the escape velocity of the black hole. Arakelian 564 has very low velocity outflows, or the absorber could even be inflowing; using the maximum outflow velocity for Phase 3 of 110\,km\,s$^{-1}$ we can calculate a minimum distance to the absorber of

\begin{equation}
R\geq \frac{2GM}{v^2_{out}},
\end{equation}
where $G$ is the gravitational constant and $M$ is the mass of the black hole ($=3.7\times 10^7\,M_{\odot}$, \nocite{matsumoto04}Matsumoto et al. 2004). This gives a value of $R\geq 25$\,pc. This is consistent with Equation~\ref{564_Rleq}, which gave a value of $R\leq$\,100\,pc for Phase 3. Assuming that this Phase must be travelling faster than the escape velocity and $R\leq$\,100\,pc, we have a minimum outflow velocity of $\sim60$\,km\,s$^{-1}$.

\subsection{Analysis of Individual Observations}
\label{564_indiv}

In order to justify coadding all five spectra we must be content that there are no significant spectral changes between the observations. Fig.~\ref{564_separate} shows that changes in the continuum and overall shape are minimal. Due to the poor signal-to-noise of many of the spectra it is not possible to fit them all separately. However, Fig.~\ref{564_separate} indicates two observations at a higher level of luminosity, and three at a lower level. Any spectral changes that are present would be expected between these two states and so the two high-state observations were added together and so were the three low-states to improve the signal-to-noise. These two spectra were then fitted with the above model, including five emission lines and three warm absorber phases. In the high-state, three warm absorber phases are found with ionization parameters log\,$\xi=-0.3\pm0.5, 1.0\pm0.4, 2.2\pm0.4$ and column densities $N_H=0.9\pm0.7, 1.4\pm1.1, 7.7\pm2.0\,\times10^{20}$\,cm$^{-2}$. The low-state warm absorbers have ionization parameters log\,$\xi=-0.5\pm0.4, 1.1\pm0.6, 2.2\pm2.1$ and column densities $N_H=1.0\pm0.9, 2.1\pm1.3, 7.2\pm1.7\,\times 10^{20}$\,cm$^{-2}$. Comparing these parameters with those in Table~\ref{564_xabs} we see that all are compatible. Therefore there are no significant spectral changes between observations and they can be coadded to produce one high signal-to-noise spectrum.

\section{Discussion}
\label{discussion}

\subsection{X-ray and UV Absorption}

\begin{figure}
\includegraphics[angle=270, scale=0.37]{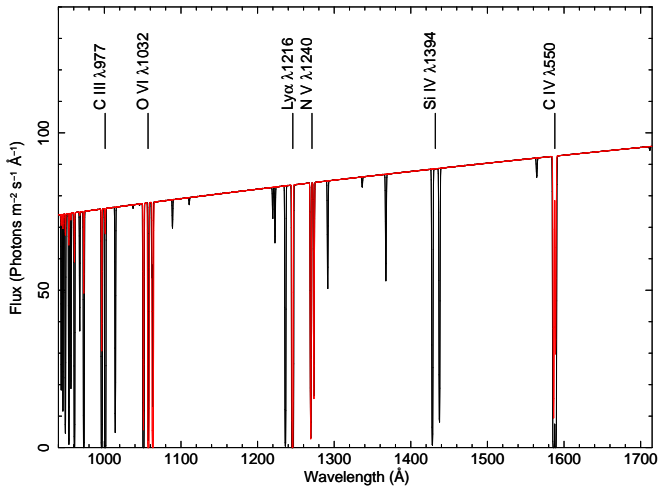}
\caption{Model graph of the expected UV spectrum, due to Phase 1 and Phase 2 absorption (Phase 1 in black, Phase 2 in red).}
\label{564_uv_model}
\end{figure}

By combining all \textit{XMM-Newton} RGS observations of Arakelian 564 we have identified three phases of photoionized, X-ray absorbing gas in its spectrum. The model used for the three phases can be seen in Fig.~\ref{564_xmodel}: Phase 1 (black) contributes absorption from very low ionization iron (Fe\,II--Fe\,IV); Phase 2 (red) contributes the mid-ionization iron M-shell UTA features along with those from ions such as N\,VI, O\,VI and C\,VI; the high ionization iron, oxygen and neon features are produced by Phase 3 (blue). The low ionization of Phases 1 and 2 means that that they will contribute significantly to the UV spectrum: this can be seen in Fig.~\ref{564_uv_model}, where strong carbon and oxygen absorption features are visible overposed on the extrapolation of the RGS continuum by the \sc xabs \rm model component in \sc spex\rm.
\newline

\begin{figure*}
\includegraphics[angle=0, scale=0.247]{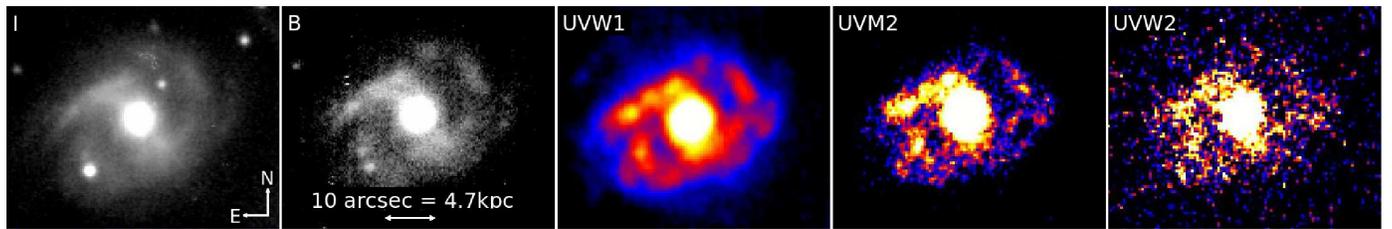}
\caption{From left to right: I-band and B-band images, from the Lick Observatory; UVW1, UVM2 and UVW2 images, from the \textit{XMM-Newton} OM, of Arakelian 564.}
\label{564_opt_uv}
\end{figure*}

\begin{figure*}
\begin{center}
\includegraphics[angle=0, scale=0.6]{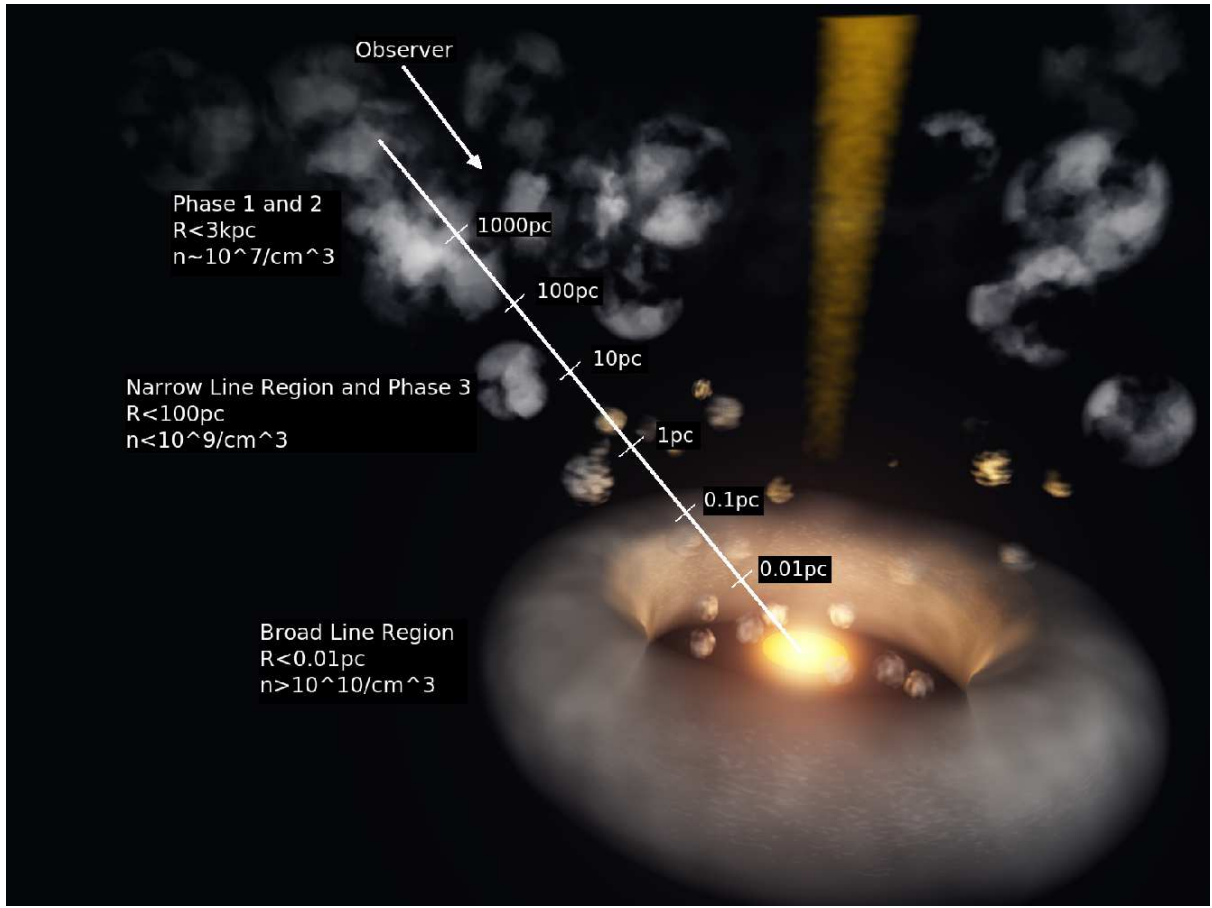}
\caption{This artist's impression of the core of Arakelian 564 shows the limits on the locations of the broad emission, narrow emission and absorption line regions.}
\label{564_cartoon}
\end{center}
\end{figure*}

Our, more detailed, findings are in reasonable agreement with~\citet{dewangan07}. Phase 3 is similar to their higher ionization warm absorber phase in ionization and column density. Their lower ionization phase lies between our Phase 1 and 2 in ionization and column density. Our Phases 2 and 3 match reasonably well the results from~\citet{matsumoto04} in ionization and velocity profile but the~\citet{matsumoto04} outflow column densities are larger (phase 1: log\,$\xi\sim1$, N$_{H}\sim10^{21}$\,cm$^{-2}$; phase 2: log\,$\xi\sim2$, N$_{H}\sim10^{21}$\,cm$^{-2}$). Our Phase 2 is also compatible with the UV absorption previously identified by~\citet{crenshaw02} and~\citet{romano02}, who found a phase with log\,$\xi\sim1.2$ and log\,$N_H\sim10^{21}$cm$^{-2}$. Therefore, it is likely that this and our Phase 2 are the same absorber. Phase 3 is too highly ionized to be detectable in the UV.
\newline

The low outflow velocities of these phases of absorbing gas are unusual. NLS1s are thought to be very luminous for their mass, and so in general would be expected to have strong, fast outflows such as those seen in NGC 4051 (\nocite{ogle04}Ogle et al. 2004), and MCG-6-30-15 (\nocite{sako03}Sako et al. 2003). On the other hand, what we see here is a relatively weak absorber with a very low outflow velocity. The low ionization phases at low velocity could be explained as interstellar gas. We have shown that Phases 1 and 2 could be up to 1000\,kpc and 10\,kpc respectively from the core; however, it is likely that these phases are somewhat closer in, as a substantial column of ionized gas would not be expected outside of the body of the host galaxy, and so we can put a limit on the distance to the absorber of the size of the galactic bulge. From examining the B-band image in Fig.~\ref{564_opt_uv} it is clear that the visible extent of the bulge is at most 3\,kpc. At this distance the gas kinematics will be dominated by the gravitational potential of the host galaxy rather than that of the black hole. Phase 3 is highly ionized and so must be part of the AGN system, and yet it has a very low flow velocity. For Phase 3 to exceed the escape velocity it must have an outflow speed in excess of 60\,km\,s$^{-1}$. If it does not, we are looking at either a stationary warm absorber or one with a transverse movement with respect to our line of sight.  
\newline

By fitting the data with a continuous distribution of photoionized gas, we have shown that three separate phases are likely to exist, rather than a continuous outflow with changing ionization and column density. We can also say that the three phases are unlikely to co-exist in the same region of the AGN system, since they are not in pressure equilibrium. If they did exist within the same region the pressure difference would force the higher pressure region (Phase 1) into the lower pressure regions (Phases 2 and 3) and what we are observing would not be a stable condition. The phases being separate supports the idea of the low ionization phases being interstellar gas in the host galaxy and Phase 3 being part of the inner AGN system, with a far higher level of ionization. We are also able to put reasonable constraints on the location of Phase 3. Our calculations in Section~\ref{sec_ion} show the distance from the central source to Phase 3 to be $25<R<100\,$pc. Using the expression for the distance to the inner edge of the torus, $\sim1\times L^{0.5}_{ion,44}$\,pc (\nocite{krolik01}Krolik \& Kriss 2001), we get $r_{torus}\sim3.9$\,pc. Because of their expected contribution to the UV spectrum Phases 1 and 2 are likely to lie outside the majority of the NLR, putting them well outside the base of the dusty torus, eliminating an accretion disk origin; however, as we have seen, Phase 3 could be much closer in and could even be connected to the NLR itself. 

\subsection{Emission Features and Connections with the Absorber}

It is clear that the X-ray absorption does not originate in the same region as the majority of the emission lines. The emission lines display higher outflow velocities (see Tables~\ref{564_emis} and~\ref{564_xabs}), suggesting an origin closer to the centre of the system. This agrees with the UV data (\nocite{crenshaw02}Crenshaw et al. 2002,~\nocite{romano02}Romano et al. 2002) in which the absorption lines eat into the emission lines, implying that the absorption occurs further out from the nucleus than the emission. It is important to note that the velocity profiles of the emission lines may be altered due to the presence of absorption lines close by, as can be seen in Fig.~\ref{564_parts}. From Fig.~\ref{564_parts} it is clear, particularly in the O\,VIII Ly$\alpha$ profile, that the absorption lines lie to the red of the emission lines. Therefore, the emission line producing gas must be outflowing faster than the absorption line region. The O\,VIII Ly$\alpha$, N\,VI(i) and N\,VII Ly$\alpha$ emission line flow velocities are consistent, within errors, indicating a likely common origin. The C\,VI Ly$\alpha$ emission line appears to have a far higher outflow velocity than the others. This could be an artifact of absorption feature contamination, being the least significant of the fitted emission lines.
\newline

O\,VIII Ly$\alpha$, N\,VI(i) and N\,VII Ly$\alpha$ have velocity widths of $\sim1000$\,km\,s$^{-1}$ making their likely origin the broad line region close in to the centre of the system. This would be expected to be a region of high density, as is derived for the optical and UV broad line region ($n_e\sim10^{11}$cm$^{-3}$) by~\citet{romano04}. The presence of an N\,VI(i) emission line without any clear identification of a forbidden line implies a region of high density. Following the analysis of photoionized plasmas by~\citet{porquet00} we define the ratio of the forbidden line $(z)$ to the intercombination lines $(x+y)$ as $R=z/(x+y)$. From the data we find a value of $R<<1$ for the N\,VI producing region. From the~\citet{porquet00} and~\citet{godet04} plots of the ratio, $R$, against density, $n_e$, we see that $R<<1$ for N\,VI implies a density of $n_e>>3\times 10^{10}$\,cm$^{-3}$. This is consistent with the BLR density of $\sim10^{11}$cm$^{-3}$ calculated by~\citet{romano04}. 
\newline

The O\,VII(f) emission line is seen to have a very low outflow velocity compared to the other soft X-ray emission lines identified here. This could be explained by it being produced in a different region of the AGN: the narrow line region. The measured velocities are in agreement with the optical and UV narrow line region values (FWHM$\leq1000$\,km\,s$^{-1}$, blueshifted by 400--1200\,km\,s$^{-1}$), as measured by~\citet{crenshaw02},~\citet{romano02} and~\citet{contini03}. Interestingly the profile of the O\,VII(f) emission line is also consistent in velocity and ionization with Phase 2 or 3 absorption. Therefore, it is possible that a component of the absorber and this emitter have the same origin. The O\,VII(f) emission line is density-sensitive, so we again use the~\citet{porquet00} analysis. We find a value of $R>3$ for the O\,VII emission triplet. This implies a density of $n_e<10^9$\,cm$^{-3}$, in agreement with NLR conditions and not those of the BLR, adding credibility to the theory that the O\,VII(f) line originates in a different region to the other identified soft X-ray emission lines.

\subsection{Comparison with warm absorbers in other NLS1s}

We have compared these results with a number of other NLS1s. NGC 4051 displays a similar continuum shape to Arakelian 564, with a steep soft X-ray continuum fitted with a power law of slope $\Gamma$=2.2 (\nocite{ogle04}Ogle et al. 2004). These authors note the existence of a broad C\,VI Ly$\alpha$ emission line, the core of which is found to have a FWHM velocity of 1200\,km\,s$^{-1}$, from a region of density $n_e=1\times 10^{10}$\,cm$^{-3}$ at $\sim2\times 10^{-3}$\,pc from the central source, with a possible outflow in the outer BLR; this FWHM is compatible with our `broad' emission lines, though the density is slightly lower. This line also has a broad component implying far higher FWHM velocities and densities, which is not seen in the emission line profiles of Arakelian 564. Also identified, by~\citet{ogle04}, are a number of narrow emission lines, due to N\,VI, O\,VII, Ne\,IX and Si\,XIII with densities $n_e=10^7-10^8$\,cm$^{-3}$, consistent with the density we find for the narrow line region in Arakelian 564. There is a large range in ionization of the absorbers of NGC 4051 which is inconsistent with being in pressure equilibrium, as also seen here.~\citet{ogle04} indicate that the narrow lines and an ionized absorber in NGC 4051 may arise in the same region.  
\newline

The NLS1s MCG-6-30-15 and Markarian 766 also show strong emission and narrow absorption lines in their RGS soft X-ray spectra (\nocite{branduardi01}Branduardi-Raymont et al. 2001). Their fits include steep power laws ($\Gamma$=2.14 and 2.53 respectively) and relativistic emission lines from O\,VIII, N\,VII and C\,VI, as well as narrow absorption lines, in order to explain the peculiar shape of the spectrum.~\citet{lee03} have also analysed \textit{Chandra} HETGS data of MCG-6-30-15 and favour a fit of the spectrum using purely a dusty warm absorber model. The analysis of the object is still the subject of much controversy;~\citet{sako03} have shown that the~\citet{lee03} model is a poor fit outside of the HETG range and have also made a detailed analysis of the warm absorber present in the two sources.~\citet{sako03} identify two warm absorber phases in MCG-6-30-15 with column densities of $\sim2\times10^{21}\,$cm$^{-2}$ and outflow velocities of $\sim-150$ and $\sim-1900$\,km\,s$^{-1}$ respectively.~\citet{mason03} use a longer RGS exposure to fit the absorption lines of Markarian 766 individually with column densities of $3.4\times10^{16}$\,cm$^{-2}$ for C\,VI and $2.2\times10^{16}$\,cm$^{-2}$ for O\,VII with a velocity width of $\sim100$\,km\,s$^{-1}$ and a small outflow velocity $\leq100$\,km\,s$^{-1}$
\newline

Two warm absorber phases are also found in the spectrum of IZw1 by~\citet{costantini07}. The two absorbers, which have ionization parameters of log\,$\xi\sim0$ and log\,$\xi\sim2.5$, have similar column densities of $\sim1.3\times10^{21}$\,cm$^{-2}$. The lower ionization phase has an outflow velocity of $\sim1700$\,km\,s$^{-1}$, and the higher of $\leq800$\,km\,s$^{-1}$; implying far higher outflow velocities than those in the Arakelian 564 absorbers. The warm absorber phases in IZw1 are similar in ionization to our Phases 1 and 3; the lower ionization phase of the two is contributing significantly to the UV spectrum, as seen in Arakelian 564. The shape of the thermal stability graph in IZw1 is also very similar to that seen here (Fig.~\ref{564_preseq}) in that there is no unstable region.~\citet{costantini07} suggest that this could be due to the steep soft X-ray spectra displayed by many NLS1s. They also find that, like ours, their warm absorber phases are not in pressure equilibrium and so would require, for example, magnetic confinement in order to co-exist. 
\newline

Markarian 359 is a NLS1 which, like Arakelian 564, has a very weak X-ray absorber.~\citet{obrien01} detect evidence for a weak Iron M-shell UTA, but no other significant absorption features. Also identified in the RGS spectrum, are three narrow emission lines at 13.6, 22.0 and 18.9\AA, corresponding to the Ne\,IX and O\,VII triplets and the O\,VIII Ly$\alpha$ emission line. The signal-to-noise of the spectrum is such that accurate velocity values cannot be well constrained. Ton S180 has a weak warm absorber, though with a larger column than that identified in Arakelian 564.~\citet{rozanska04} identify a number of narrow absorption lines with a minimum column density of $5.1\times 10^{21}$\,cm$^{-2}$. There is also emission from C\,VI, O\,VIII and Ne\,X. Unfortunately, due to the weakness of these features, very little information can be determined in order to make a comparison with Arakelian 564.
\newline

By comparing Arakelian 564 with other NLS1s we can see that it is an unusual object, but at least two other galaxies, Markarian 359 and Ton S180, may have similar warm absorber properties. The warm absorber in Arakelian 564 is very weak and has a very low outflow velocity. Unfortunately the available data for Markarian 359 and Ton S180 are not of sufficient quality to determine outflow velocities; the only NLS1 warm absorber that we have found in the literature, with well defined parameters and such low velocities as Arakelian 564, is that in Markarian 766 (e.g.~\nocite{mason03}Mason et al. 2003). It is also unusual to find strong, broad, blueshifted emission lines in the soft X-ray spectrum, with similar profiles as seen in the UV lines, raising questions about what happens to this gas and how it may be connected to the narrow emission and absorption line regions. 

\subsection{Physical Structure}

\begin{figure*}
\begin{center}
\includegraphics[angle=0, scale=0.3]{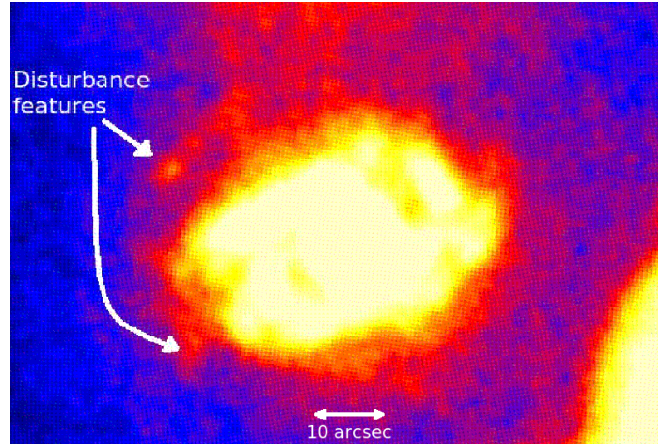}
\caption{UVW1 image of Arakelian 564 with the contrast set such that disturbance features are enhanced.}
\label{564_uvw1}
\end{center}
\end{figure*}

We can see from the images in Fig.~\ref{564_opt_uv} that the host galaxy is of SBb type (e.g.~\nocite{petrov99}Petrov et al. 1999). The UV images show that the galaxy has some unusual characteristics, such as its square shape. Where the I-band and B-band images show evidence of a fairly typical SBb galaxy, the UV images show some features that do not conform to the spiral structure of the galaxy. These features are most noticeable in the north-east and south-east corners of the UVW1 image as indicated in Fig.~\ref{564_uvw1}. They are a possible indication of some disturbance to the galaxy in the past, producing tidal streams. It is possible that the disturbed nature of the host galaxy would be the cause of some of the absorption and extinction features observed in Arakelian 564.
\newline

We can estimate the viewing angle of the host galaxy by measuring its ellipticity. This is done by measuring the major and minor axis of a number of isophotes fitted to the I-band image of the galaxy. From averaging these values we find an ellipticity value, defined as $e=1-b/a$ (a = major axis, b=minor axis), of $e=0.184$; this compares well with the measured value of 0.198 from~\citet{schmitt00}. Using our value of $e=0.184$, and assuming an intrinsically circular galaxy disk, we calculate an inclination of $35\pm10^{\circ}$ (errors calculated at 90\% confidence from measuring the ellipticity of the galaxy at a number of different isophotes) in agreement with the $37^{\circ}$ value calculated from the~\citet{schmitt00} ellipticity. If the plane of the AGN accretion disk was aligned with the plane of the host galaxy then we would expect to find that the position angle (P.A.) of the radio jet in this object lies along the same axis as the host galaxy minor axis. \citet{schmitt01} find that the radio structure of Arakelian 564 lies in a north-south direction (P.A. $\sim6^{\circ}$). From Fig.~\ref{564_opt_uv} we can see that the P.A. of the minor axis of the host galaxy is $\sim50^{\circ}$, and so the two are not aligned.  
\newline

From our analysis and previous studies of the UV and optical spectra it is clear that the BLR in Arakelian 564 has a reasonably large outflow velocity of $\sim1000$\,km\,s$^{-1}$, the NLR a slightly lower one at a few hundred km\,s$^{-1}$ and the warm absorber is outflowing at a maximum speed of $\sim100$\,km\,s$^{-1}$. All three components display a very similar range of ionization.~\citet{crenshaw02} state that the warm absorber contains a large amount of dust ($E(B-V)=0.14$); using the standard Galactic gas-to-dust ratio this implies gas with a column density $\geq8\times 10^{20}$\,cm$^{-2}$. At this gas-to-dust ratio, all three of our absorber phases would be needed to contain this dust and therefore the majority of this material must lie outside of the NLR in order to significantly redden the emission lines. Warm absorbers, in general, do not contain large amounts of dust, making this object quite unusual.
\newline 

We can calculate mass outflow rates for the different regions of the AGN. For the BLR we first need to estimate the filling factor using the equation

\begin{equation}
Filling factor = \frac{3L(H\alpha)}{4\pi n_H^2\alpha_{H\alpha}^{eff}hv_{H\alpha}r^3}
\end{equation}
from~\citet{devereux07}, where $L(H\alpha)=5.8\times 10^{40}$\,erg\,s$^{-1}$ is the luminosity of the $H\alpha$ line, $n_H=10^{11}$\,cm$^{-3}$ is the density from~\citet{romano04}, $\alpha_{H\alpha}^{eff}=9\times 10^{-14}$\,cm$^{-3}$\,s$^{-1}$ is the effective recombination coefficient as defined in~\citet{osterbrock06}, and $r$ is the distance of the BLR from the centre of the system.
\newline

This gives a filling factor of approximately $3\times 10^{-7}$. Using this we can estimate a mass outflow rate of the BLR as $\sim 1.3\times 10^{-4}M_{\odot}$yr$^{-1}$. Following the method of~\citet{blustin05}, we calculate the total mass outflow rate of the absorbing regions as $\sim$ 0.6, 0.2 and 0.1$M_{\odot}$\,yr$^{-1}$ for Phases 1, 2 and 3 respectively. This is significantly larger than the BLR mass outflow rate, which could be explained by the BLR sweeping material off the torus and creating the dusty warm absorber. Mass loading of the gas would also slow it down, in order to conserve momentum, to the required speeds observed in the absorber.
\newline

Combining this evidence suggests that the BLR, NLR and absorber may have a common origin. The outflowing BLR becomes more diffuse with greater distance from the central system, keeping the ionization of the gas approximately the same. Part of this outflow sweeps material off the edge of the torus, slowing the outflow and increasing its dust content. This then becomes the less clumpy NLR and the diffuse warm absorber, as is shown in Fig.~\ref{564_cartoon}.  
\newline

Alternatively one or more of the phases of ionized absorber could be up to a few kpc from the core. This could be explained by the observation that the majority of NLS1s contain bars (\nocite{crenshaw03a}Crenshaw et al. 2003), which enable the transport of dust and gas along the leading edge of the bar (\nocite{sparke00}Sparke \& Gallagher 2000). This material infalls towards the centre until it reaches a stable circular orbit around the galactic bulge. This then either stays in orbit or infalls in a circular or barlike motion. This ring of gas and dust could be the source of our low ionization absorber phases, explaining low flow velocity. Also, disturbance to this galaxy, of which we have tentative evidence, could have caused gas and dust to be repositioned throughout the system, giving rise to dust and gas lanes within our line of sight.

\section{Conclusion}

We present a detailed analysis of the photoionized, X-ray absorbing and emitting gas present in the NLS1 galaxy Arakelian 564. The high-resolution X-ray spectra are from five observations with \textit{XMM-Newton}'s Reflection Grating Spectrometer.
\newline

The absorption profile is fitted with three phases of X-ray absorbing gas with average RMS velocity $\sim60$\,km\,s$^{-1}$ and flow velocity $\sim-10$\,km\,s$^{-1}$. Each phase has a different ionization (log $\xi=-0.86, 0.87, 2.56$) and column density ($N_H=0.89\times10^{20}, 2.41\times10^{20}, 6.03\times10^{20}$\,cm$^{-2}$ respectively). Note that one continuous distribution of gas is unlikely, but not ruled out. The low flow velocity of the warm absorber phases can be partially explained by assuming the low ionization phases to be interstellar gas. The lower ionization phases are expected to contribute significantly to the UV spectrum, implying that the UV and X-ray absorbers are connected.
\newline

The best fit to the RGS data requires five significant emission lines due to O\,VIII Ly$\alpha$ (18.9\AA), O\,VII(f) (22.1\AA), N\,VI(i) (29.0\AA), N\,VII Ly$\alpha$ (24.7\AA) and C\,VI Ly$\alpha$ (33.5\AA). By examining the outflow velocities and density restrictions on these emission lines we have shown that they originate in two separate regions, the BLR and the NLR. It is possible that these regions and the absorber are connected. The outflowing broad line region expands as it travels out from the core, maintaining a similar ionization level. This gas then picks up dust from the torus, and slows down. As the gas continues to expand it then displays the characteristics of the NLR and the diffuse, dusty warm absorber. UV observations suggest a disturbed host galaxy. A tidal stream or dust and gas accumulating in the inner regions of the host galaxy offer an alternative origin for the low ionization absorber.

\acknowledgements

The work shown here is based on observations obtained with \textit{XMM-Newton}, an ESA science mission with instruments and contributions directly funded by ESA Member states and the USA(NASA). The authors acknowledge the assistance from G.A.Collinson with the 3D modelling used in this paper. R.A.N.S. acknowledges the support of a STFC studentship.

\bibliography{/disk/xray2/rans/thesis/bibliography}

\begin{thebibliography}{49}
\expandafter\ifx\csname natexlab\endcsname\relax\def\natexlab#1{#1}\fi

\bibitem[{Blustin {et~al.}(2005)Blustin, Page, Fuerst, Branduardi-Raymont, \&
  Ashton}]{blustin05}
Blustin, A.~J., Page, M.~J., Fuerst, S.~V., Branduardi-Raymont, G., \& Ashton,
  C.~E. 2005, A\&A, 431, 111

\bibitem[{{Boller} {et~al.}(1996){Boller}, {Brandt}, \& {Fink}}]{boller96}
{Boller}, T., {Brandt}, W.~N., \& {Fink}, H. 1996, A\&A, 305, 53

\bibitem[{{Boroson}(2002)}]{boroson02}
{Boroson}, T.~A. 2002, ApJ, 565, 78

\bibitem[{{Branduardi-Raymont} {et~al.}(2001){Branduardi-Raymont}, {Sako},
  {Kahn}, {Brinkman}, {Kaastra}, \& {Page}}]{branduardi01}
{Branduardi-Raymont}, G., {Sako}, M., {Kahn}, S.~M., {et~al.} 2001, A\&A, 365,
  L140

\bibitem[{{Contini} {et~al.}(2003){Contini}, {Rodr{\'{\i}}guez-Ardila}, \&
  {Viegas}}]{contini03}
{Contini}, M., {Rodr{\'{\i}}guez-Ardila}, A., \& {Viegas}, S.~M. 2003, A\&A,
  408, 101

\bibitem[{{Costantini} {et~al.}(2007){Costantini}, {Gallo}, {Brandt}, {Fabian},
  \& {Boller}}]{costantini07}
{Costantini}, E., {Gallo}, L.~C., {Brandt}, W.~N., {Fabian}, A.~C., \&
  {Boller}, T. 2007, MNRAS, 378, 873

\bibitem[{{Crenshaw} {et~al.}(2003){Crenshaw}, {Kraemer}, \&
  {Gabel}}]{crenshaw03a}
{Crenshaw}, D.~M., {Kraemer}, S.~B., \& {Gabel}, J.~R. 2003, AJ, 126, 1690

\bibitem[{{Crenshaw} {et~al.}(2002){Crenshaw}, {Kraemer}, {Turner}, {Collier},
  {Peterson}, {Brandt}, {Clavel}, {George}, {Horne}, {Kriss}, {Mathur},
  {Netzer}, {Pogge}, {Pounds}, {Romano}, {Shemmer}, \&
  {Wamsteker}}]{crenshaw02}
{Crenshaw}, D.~M., {Kraemer}, S.~B., {Turner}, T.~J., {et~al.} 2002, ApJ, 566,
  187

\bibitem[{{den Herder} {et~al.}(2001){den Herder}, {Brinkman}, {Kahn},
  {Branduardi-Raymont}, {Thomsen}, {Aarts}, {Audard}, {Bixler}, {den Boggende},
  {Cottam}, {Decker}, {Dubbeldam}, {Erd}, {Goulooze}, {G{\"u}del}, {Guttridge},
  {Hailey}, {Janabi}, {Kaastra}, {de Korte}, {van Leeuwen}, {Mauche},
  {McCalden}, {Mewe}, {Naber}, {Paerels}, {Peterson}, {Rasmussen}, {Rees},
  {Sakelliou}, {Sako}, {Spodek}, {Stern}, {Tamura}, {Tandy}, {de Vries},
  {Welch}, \& {Zehnder}}]{denherder01}
{den Herder}, J.~W., {Brinkman}, A.~C., {Kahn}, S.~M., {et~al.} 2001, A\&A,
  365, L7

\bibitem[{{Devereux} \& {Shearer}(2007)}]{devereux07}
{Devereux}, N. \& {Shearer}, A. 2007, ApJ, 671, 118

\bibitem[{{Dewangan} {et~al.}(2007){Dewangan}, {Griffiths}, {Dasgupta}, \&
  {Rao}}]{dewangan07}
{Dewangan}, G.~C., {Griffiths}, R.~E., {Dasgupta}, S., \& {Rao}, A.~R. 2007,
  ArXiv e-prints, 709

\bibitem[{{George} {et~al.}(1998){George}, {Turner}, {Netzer}, {Nandra},
  {Mushotzky}, \& {Yaqoob}}]{george98}
{George}, I.~M., {Turner}, T.~J., {Netzer}, H., {et~al.} 1998, ApJS, 114, 73

\bibitem[{{Godet} {et~al.}(2004){Godet}, {Collin}, \& {Dumont}}]{godet04}
{Godet}, O., {Collin}, S., \& {Dumont}, A.-M. 2004, A\&A, 426, 767

\bibitem[{{Grupe}(1996)}]{grupe96}
{Grupe}, D. 1996, PhD thesis, PhD thesis.~Univ.~G{\"o}ttingen , (1996)

\bibitem[{{Hasinger}(1997)}]{hasinger97}
{Hasinger}, G. 1997, in X-Ray Imaging and Spectroscopy of Cosmic Hot Plasmas,
  ed. F.~{Makino} \& K.~{Mitsuda}, 263--+

\bibitem[{{Huchra} {et~al.}(1999){Huchra}, {Vogeley}, \& {Geller}}]{huchra99}
{Huchra}, J.~P., {Vogeley}, M.~S., \& {Geller}, M.~J. 1999, ApJS, 121, 287

\bibitem[{{Kalberla} {et~al.}(2005){Kalberla}, {Burton}, {Hartmann}, {Arnal},
  {Bajaja}, {Morras}, \& {P{\"o}ppel}}]{kalberla05}
{Kalberla}, P.~M.~W., {Burton}, W.~B., {Hartmann}, D., {et~al.} 2005, A\&A,
  440, 775

\bibitem[{{Komossa}(2000)}]{komossa00}
{Komossa}, S. 2000, New Astronomy Review, 44, 483

\bibitem[{{Krolik} \& {Kriss}(2001)}]{krolik01}
{Krolik}, J.~H. \& {Kriss}, G.~A. 2001, ApJ, 561, 684

\bibitem[{{Krolik} {et~al.}(1981){Krolik}, {McKee}, \& {Tarter}}]{krolik81}
{Krolik}, J.~H., {McKee}, C.~F., \& {Tarter}, C.~B. 1981, ApJ, 249, 422

\bibitem[{{Lampton} {et~al.}(1976){Lampton}, {Margon}, \& {Bowyer}}]{lampton76}
{Lampton}, M., {Margon}, B., \& {Bowyer}, S. 1976, ApJ, 208, 177

\bibitem[{{Laor} {et~al.}(1994){Laor}, {Fiore}, {Elvis}, {Wilkes}, \&
  {McDowell}}]{laor94}
{Laor}, A., {Fiore}, F., {Elvis}, M., {Wilkes}, B.~J., \& {McDowell}, J.~C.
  1994, ApJ, 435, 611

\bibitem[{{Lee} {et~al.}(2003){Lee}, {Canizares}, {Fang}, {Kallman}, {Fabian},
  {Turner}, \& {Marshall}}]{lee03}
{Lee}, J.~C., {Canizares}, C.~R., {Fang}, T., {et~al.} 2003, in Bulletin of the
  American Astronomical Society, Vol.~35, Bulletin of the American Astronomical
  Society, 635--+

\bibitem[{{Leighly}(1999)}]{leighly99}
{Leighly}, K.~M. 1999, ApJS, 125, 317

\bibitem[{{Mason} {et~al.}(2003){Mason}, {Branduardi-Raymont}, {Ogle}, {Page},
  {Puchnarewicz}, {Behar}, {C{\'o}rdova}, {Davis}, {Maraschi}, {McHardy},
  {O'Brien}, {Priedhorsky}, \& {Sasseen}}]{mason03}
{Mason}, K.~O., {Branduardi-Raymont}, G., {Ogle}, P.~M., {et~al.} 2003, ApJ,
  582, 95

\bibitem[{{Mason} {et~al.}(2001){Mason}, {Breeveld}, {Much}, {Carter},
  {Cordova}, {Cropper}, {Fordham}, {Huckle}, {Ho}, {Kawakami}, {Kennea},
  {Kennedy}, {Mittaz}, {Pandel}, {Priedhorsky}, {Sasseen}, {Shirey}, {Smith},
  \& {Vreux}}]{mason01}
{Mason}, K.~O., {Breeveld}, A., {Much}, R., {et~al.} 2001, A\&A, 365, L36

\bibitem[{{Matsumoto} {et~al.}(2004){Matsumoto}, {Leighly}, \&
  {Marshall}}]{matsumoto04}
{Matsumoto}, C., {Leighly}, K.~M., \& {Marshall}, H.~L. 2004, ApJ, 603, 456

\bibitem[{{O'Brien} {et~al.}(2001){O'Brien}, {Page}, {Reeves}, {Pounds},
  {Turner}, \& {Puchnarewicz}}]{obrien01}
{O'Brien}, P.~T., {Page}, K., {Reeves}, J.~N., {et~al.} 2001, MNRAS, 327, L37

\bibitem[{{Ogle} {et~al.}(2004){Ogle}, {Mason}, {Page}, {Salvi}, {Cordova},
  {McHardy}, \& {Priedhorsky}}]{ogle04}
{Ogle}, P.~M., {Mason}, K.~O., {Page}, M.~J., {et~al.} 2004, ApJ, 606, 151

\bibitem[{{Osterbrock} \& {Ferland}(2006)}]{osterbrock06}
{Osterbrock}, D.~E. \& {Ferland}, G.~J. 2006, {Astrophysics of gaseous nebulae
  and active galactic nuclei} (Astrophysics of gaseous nebulae and active
  galactic nuclei, 2nd.~ed.~by D.E.~Osterbrock and G.J.~Ferland.~Sausalito, CA:
  University Science Books, 2006)

\bibitem[{{Osterbrock} \& {Pogge}(1985)}]{osterbrock85}
{Osterbrock}, D.~E. \& {Pogge}, R.~W. 1985, ApJ, 297, 166

\bibitem[{{Page} {et~al.}(2003){Page}, {Soria}, {Wu}, {Mason}, {Cordova}, \&
  {Priedhorsky}}]{page03a}
{Page}, M.~J., {Soria}, R., {Wu}, K., {et~al.} 2003, MNRAS, 345, 639

\bibitem[{{Papadakis} {et~al.}(2007){Papadakis}, {Brinkmann}, {Page},
  {McHardy}, \& {Uttley}}]{papadakis06}
{Papadakis}, I.~E., {Brinkmann}, W., {Page}, M.~J., {McHardy}, I., \& {Uttley},
  P. 2007, A\&A, 461, 931

\bibitem[{{Petrov} {et~al.}(1999){Petrov}, {Slavcheva}, {Bachev}, \&
  {Mihov}}]{petrov99}
{Petrov}, G., {Slavcheva}, L., {Bachev}, R., \& {Mihov}, B. 1999, in IAU
  Symposium, Vol. 194, Activity in Galaxies and Related Phenomena, ed.
  Y.~{Terzian}, E.~{Khachikian}, \& D.~{Weedman}, 84--+

\bibitem[{{Porquet} \& {Dubau}(2000)}]{porquet00}
{Porquet}, D. \& {Dubau}, J. 2000, A\&AS, 143, 495

\bibitem[{{Pounds} {et~al.}(1995){Pounds}, {Done}, \& {Osborne}}]{pounds95}
{Pounds}, K.~A., {Done}, C., \& {Osborne}, J.~P. 1995, MNRAS, 277, L5

\bibitem[{{Romano} {et~al.}(2002){Romano}, {Mathur}, {Pogge}, {Peterson}, \&
  {Kuraszkiewicz}}]{romano02}
{Romano}, P., {Mathur}, S., {Pogge}, R.~W., {Peterson}, B.~M., \&
  {Kuraszkiewicz}, J. 2002, ApJ, 578, 64

\bibitem[{{Romano} {et~al.}(2004){Romano}, {Mathur}, {Turner}, {Kraemer},
  {Crenshaw}, {Peterson}, {Pogge}, {Brandt}, {George}, {Horne}, {Kriss},
  {Netzer}, {Shemmer}, \& {Wamsteker}}]{romano04}
{Romano}, P., {Mathur}, S., {Turner}, T.~J., {et~al.} 2004, ApJ, 602, 635

\bibitem[{{R{\'o}{\.z}a{\'n}ska} {et~al.}(2004){R{\'o}{\.z}a{\'n}ska},
  {Czerny}, {Siemiginowska}, {Dumont}, \& {Kawaguchi}}]{rozanska04}
{R{\'o}{\.z}a{\'n}ska}, A., {Czerny}, B., {Siemiginowska}, A., {Dumont}, A.-M.,
  \& {Kawaguchi}, T. 2004, ApJ, 600, 96

\bibitem[{{Sako} {et~al.}(2003){Sako}, {Kahn}, {Branduardi-Raymont}, {Kaastra},
  {Brinkman}, {Page}, {Behar}, {Paerels}, {Kinkhabwala}, {Liedahl}, \&
  {Herder}}]{sako03}
{Sako}, M., {Kahn}, S.~M., {Branduardi-Raymont}, G., {et~al.} 2003, ApJ, 596,
  114

\bibitem[{{Schmitt} \& {Kinney}(2000)}]{schmitt00}
{Schmitt}, H.~R. \& {Kinney}, A.~L. 2000, ApJS, 128, 479

\bibitem[{{Schmitt} {et~al.}(2001){Schmitt}, {Ulvestad}, {Antonucci}, \&
  {Kinney}}]{schmitt01}
{Schmitt}, H.~R., {Ulvestad}, J.~S., {Antonucci}, R.~R.~J., \& {Kinney}, A.~L.
  2001, ApJS, 132, 199

\bibitem[{{Shemmer} {et~al.}(2001){Shemmer}, {Romano}, {Bertram}, {Brinkmann},
  {Collier}, {Crowley}, {Detsis}, {Filippenko}, {Gaskell}, {George}, {Gliozzi},
  {Hiller}, {Jewell}, {Kaspi}, {Klimek}, {Lannon}, {Li}, {Martini}, {Mathur},
  {Negoro}, {Netzer}, {Papadakis}, {Papamastorakis}, {Peterson}, {Peterson},
  {Pogge}, {Pronik}, {Rumstay}, {Sergeev}, {Sergeeva}, {Stirpe}, {Taylor},
  {Treffers}, {Turner}, {Uttley}, {Vestergaard}, {von Braun}, {Wagner}, \&
  {Zheng}}]{shemmer01}
{Shemmer}, O., {Romano}, P., {Bertram}, R., {et~al.} 2001, ApJ, 561, 162

\bibitem[{{Sparke} \& {Gallagher}(2000)}]{sparke00}
{Sparke}, L.~S. \& {Gallagher}, III, J.~S. 2000, {Galaxies in the universe : an
  introduction} (Galaxies in the Universe, by Linda S.~Sparke and John
  S.~Gallagher, III, pp.~416.~ISBN 0521592410.~Cambridge, UK: Cambridge
  University Press, September 2000.)

\bibitem[{{Tarter} {et~al.}(1969){Tarter}, {Tucker}, \& {Salpeter}}]{tarter69}
{Tarter}, C.~B., {Tucker}, W.~H., \& {Salpeter}, E.~E. 1969, ApJ, 156, 943

\bibitem[{{Turner} {et~al.}(2001){Turner}, {Romano}, {George}, {Edelson},
  {Collier}, {Mathur}, \& {Peterson}}]{turner01}
{Turner}, T.~J., {Romano}, P., {George}, I.~M., {et~al.} 2001, ApJ, 561, 131

\bibitem[{{Vaughan} {et~al.}(1999){Vaughan}, {Pounds}, {Reeves}, {Warwick}, \&
  {Edelson}}]{vaughan99}
{Vaughan}, S., {Pounds}, K.~A., {Reeves}, J., {Warwick}, R., \& {Edelson}, R.
  1999, MNRAS, 308, L34

\bibitem[{{Verner} {et~al.}(1996){Verner}, {Verner}, \& {Ferland}}]{verner96}
{Verner}, D.~A., {Verner}, E.~M., \& {Ferland}, G.~J. 1996, Atomic Data and
  Nuclear Data Tables, 64, 1

\bibitem[{{Vignali} {et~al.}(2004){Vignali}, {Brandt}, {Boller}, {Fabian}, \&
  {Vaughan}}]{vignali04}
{Vignali}, C., {Brandt}, W.~N., {Boller}, T., {Fabian}, A.~C., \& {Vaughan}, S.
  2004, MNRAS, 347, 854

\end{thebibliography}
\end{document}